# Nanofluid under Uniform Transverse Magnetic Field with a Chemical Reaction past a Stretching Sheet

Nalini S Patil[1], Vishwambhar S. Patil [2] and J.N.Salunke[3]

[1]Department of Mathematics, Pratap College, Amalner-425401, India
[2]Department of Mathematics, Government College of Engineering, Karad-415124, India
[3]School of Mathematical Sciences, Swami Ramanand Marathawada University, Nanded, India

***Abstract*** — *An analysis is carried out to examine the effects of the steady flow of an electrically conducting viscous incompressible nanofluid in the existence of a uniform transverse magnetic field with chemical reaction over a stretching sheet considering suction or injection. The present model is demonstrated experimentally to reveal the effects of Thermophoresis and Brownian motion. Similarity solutions are investigated for the governing equations and solved numerically by using a shooting technique with fourth-order Runge-Kutta integration scheme. The impact of associated parameters on velocity profile, concentration profile and temperature profile is plotted graphically. Also the impact of local skin friction coefficient, the reduced Nusselt number and the reduced Sherwood number are discussed. The experimental setup revealed good agreement of the results with the literature.*

## I. INTRODUCTION

The most of the process in the industry includes a new strategy for the heat transfer such as in flowing of fluid in the laminar or turbulent regime. Most of these applications are intended to reduce the thermal resistance of heat transfer in fluids, which results the low cost energy efficient smaller heat transfer systems. The idea of a fusion of nanoparticles with a base fluid was first introduced by Choi [3]. Nanofluid contains nanoparticles, which are produced from metals as aluminium, iron, carbides, nitrides etc. Nano fluid has produced significant interest over recent years, due to its significance in heat transfer. In general, the enhancement of thermal conductivity found within a span of 15% - 40% across base fluid and 40% of enhancement in the coefficient of heat which is presented by Yu et al [21]. Many applications in industry like power generation, chemical operations, cooling and heating processes addressed could be extended and investigated further to study heat conduction and mass diffusion. Several studies have reported the advantages of the convective flow of nanofluid such as Das et al [7] and Nield and Bejan [17] and by Buongiorno [2], Eastman et al [8], Kakac and Pramuanjaroenkij [11] and others. There are various different mechanisms for simulating this idea of convective flow of nanofluid presented by Wang and Wei [20], Nield and Kuznetsov [13-14], Khan and Pop [12], Gorla et al [9-10] and Makinde and Aziz [15]. The interdisciplinary nature of nanofluid provides ample opportunity for the researchers to undertake further investigations.

Due to remarkable features of steady magneto-hydrodynamic flow and heat transfer characteristic over a stretching sheet, it seems to be widely used in the field of metallurgy and polymer technology. As evidence these flows can be observed in many industrial processes such as melt-spinning, strips cooling in the process of drawing, etc. For this reason, researchers are investigating alternative technologies such as the problem of chemical reaction leading to heat and mass transfer that appears in many operations like drying, evaporation and the chemical processing of materials. These approaches produced impressive results and were later extended to the other areas of applications. The experiment of MHD flow of stretched vertical porous surface with chemical reaction was performed by Chamkha [4]. Later on, similar investigations were carried out by Afifi [1]. The study of natural convection in porous media past vertical surfaces with the synthetic reaction and the effects of Soret and Dufour were studied by Postelnicu [19]. Palanimani [18] examine thermal diffusivity over the MHD boundary layer flow in porous medium.

The next section deals with a set of tests sought to examine convective heat and mass transfer effects among the governing model. Further, we discuss the Thermophoresis and Brownian motion effects. Using linear transformations, the governing model with the boundary conditions is transformed into a system of ODEs. Using fourth-order Runge-Kutta integration with shooting method, the local skin friction coefficient, the reduced Nusselt and Sherwood number, the velocity, temperature and concentration profiles are studied for associated parameters. No doubt, this study would lead to many useful applications in industry.

## II. GOVERNING EQUATIONS

Consider the steady viscous and incompressible flow of Nanofluid over the stretching sheet along the plane $y = 0$. Assuming origin to be fixed the two equal and upside down forces are employed along the x-axis to generate the flow that results stretching of the surface. The homogeneous transversal magnetic field $B(x)$ enforces along the y-axis. The induced magnetic field due to the very small Reynolds number is neglected. The electric field is omitted due to emission of charges. The ambient fluid temperature $T_\infty$ and the concentration $C_\infty$ and those at the stretching surface are $T_w(x)$ and $C_w(x)$ respectively (see Nield and Kuznetsov [13-14]). Also the pressure gradient, viscous and electrical dissipation is dropped. The physical model of the flow is governed by

$$\frac{\partial u}{\partial x} + \frac{\partial v}{\partial y} = 0 \quad (1)$$

$$u\frac{\partial u}{\partial x} + v\frac{\partial u}{\partial y} = \frac{1}{\rho}\frac{\partial}{\partial y}\left(\mu\frac{\partial u}{\partial y}\right) - \frac{\sigma B^2(x)u}{\rho} \quad (2)$$

$$u\frac{\partial T}{\partial x} + v\frac{\partial T}{\partial y} = \frac{\partial}{\partial y}\left(\alpha\frac{\partial T}{\partial y}\right) + \frac{\mu}{\rho C_p}\left(\frac{\partial u}{\partial y}\right)^2 + \frac{\sigma B^2(x)u^2}{\rho C_p}$$
$$+ \tau\left[D_B \frac{\partial C}{\partial y}\frac{\partial T}{\partial y} + \frac{D_T}{T_\infty}\left(\frac{\partial T}{\partial y}\right)^2\right] \quad (3)$$

$$u\frac{\partial C}{\partial x} + v\frac{\partial C}{\partial y} = D_B \frac{\partial^2 C}{\partial y^2} + \frac{D_T}{T_\infty}\left(\frac{\partial^2 T}{\partial y^2}\right) + K_r(C - C_\infty) \quad (4)$$

Subject to the boundary conditions,
$$u(x,0) = U(x) = C_1 x^m, \quad v(x,0) = v_w(x) = C_2 x^n,$$
$$T = T_w(x) = T_\infty + C_3 x^r, \quad C = C_w(x) = C_\infty + C_4 x^r \quad (5.1)$$

$$u(x,\infty) = 0, \quad C(x,\infty) = C_\infty, \quad T(x,\infty) = T_\infty \quad (5.2)$$

where u and v are the are the velocity components along the $xy$-directions respectively, $\rho$ is the fluid density, $\mu$ is the thermal viscosity, $\sigma$ is the electrical conductivity, T is temperature of fluid inside the thermal boundary layer and $T_\infty$ is temperature of fluid in the free stream while C and $C_\infty$ are the compatible concentrations. Also $D_B$ is the Brownian constant, $D_T$ is the coefficient of Thermophoresis diffusion, $C_1, C_2, C_3, C_4$ are the constants, $U(x) = C_1 x^m$ is the speed of stretching the plate, $v_w(x) = C_2 x^n$ is transverse velocity, $B = B_0 x^s$ is the applied magnetic field, $\alpha$ is the thermal diffusivity, k is thermal conductivity and $K_r$ is the rate of chemical reaction. Introducing stream function $\psi(x,y)$ such that, $u = \frac{\partial \psi}{\partial y}$, $v = -\frac{\partial \psi}{\partial x}$ so that equation (1) satisfied identically. By introducing non-dimensional quantities $\theta = \frac{T - T_\infty}{T_w - T_\infty}$ and $\varphi = \frac{C - C_\infty}{C_w - C_\infty}$ the equations (2)-(5) along with the boundary conditions (5.1, 2) are given by

$$\psi_y \psi_{xy} - \psi_x \psi_{yy} = \frac{\mu}{\rho}\psi_{yyy} - \frac{\sigma B_0^2 x^{2s}}{\rho}\psi_y \quad (6)$$

$$\psi_y\big((T_w - T_\infty)\theta\big)_x - \psi_x\big((T_w - T_\infty)\theta\big)_y = \alpha\big((T_w - T_\infty)\theta\big)_{yy}$$
$$+ \frac{\mu}{\rho C_p}\psi_{yy}^2 + \frac{\sigma B_0^2 x^{2s}}{\rho C_p}\psi_y^2 + \quad (7)$$
$$\tau\left[D_B\big((T_w - T_\infty)\theta\big)_y\big((C_w - C_\infty)\varphi\big)_y + \frac{D_T}{T_\infty}\big(((T_w - T_\infty)\theta)_y\big)^2\right]$$

$$\psi_y\big((C_w - C_\infty)\varphi\big)_x - \psi_x\big((C_w - C_\infty)\varphi\big)_y =$$
$$D_B\big((C_w - C_\infty)\varphi\big)_{yy} + \frac{D_T}{T_\infty}\big((T_w - T_\infty)\theta\big)_{yy} \quad (8)$$
$$+ K_r \varphi(C_w - C_\infty)$$

$$\psi_y(x,0) = C_1 x^m, \quad \psi_x(x,0) = -C_2 x^n,$$
$$\theta(x,0) = 1, \quad \varphi(x,0) = 1 \quad (9.1)$$
$$\psi_y(x,\infty) = 0, \quad \theta(x,\infty) = 0, \quad \varphi(x,\infty) = 0 \quad (9.2)$$

## III. SIMILARITY ANALYSIS

Introducing a linear transformations
$$G: \begin{cases} x = A^{\alpha_1}\bar{x}; \; y = A^{\alpha_2}\bar{y}; \psi = A^{\alpha_3}\bar{\psi}; \theta = \bar{\theta}; \\ \varphi = \bar{\varphi}; \; (T - T_\infty) = A^{\alpha_4}(\bar{T} - \bar{T}_\infty); \\ (C - C_\infty) = A^{\alpha_5}(\bar{C} - \bar{C}_\infty) \end{cases} \quad (10)$$

where $\alpha_i$ $(i = 1, 2, \cdots 5)$ are constants and A is a group transformation parameter.

We seek relations between the $\alpha_i$'s such that the equations (6)-(8) along with the boundary conditions (9.1, 2) will be remains unchanged under the group of transformation (10). This give rise to

$$A^{2\alpha_3-2\alpha_2-\alpha_1}\bar{\psi}_{\bar{y}}\bar{\psi}_{\bar{x}\,\bar{y}} - A^{2\alpha_3-2\alpha_2-\alpha_1}\bar{\psi}_{\bar{x}}\bar{\psi}_{\bar{y}\,\bar{y}} =$$
$$A^{\alpha_3-3\alpha_2}\frac{\mu}{\rho}\bar{\psi}_{\bar{y}\,\bar{y}\,\bar{y}} - A^{2\alpha_1 s+\alpha_3-\alpha_2}\frac{\sigma B_0^2 \bar{x}^{2s}}{\rho}\bar{\psi}_{\bar{y}} \quad (11)$$

$$A^{\alpha_4+\alpha_3-\alpha_2-\alpha_1}\bar{\psi}_{\bar{y}}\left((\bar{T}_w-\bar{T}_\infty)\bar{\theta}\right)_{\bar{x}} - A^{\alpha_4+\alpha_3-\alpha_2-\alpha_1}$$
$$\bar{\psi}_{\bar{x}}\left((\bar{T}_w-\bar{T}_\infty)\bar{\theta}\right)_{\bar{y}} = A^{\alpha_4-2\alpha_2}\alpha\left((\bar{T}_w-\bar{T}_\infty)\bar{\theta}\right)_{\bar{y}\,\bar{y}}$$
$$+ A^{2\alpha_3-4\alpha_2}\frac{\mu}{\rho C_p}\bar{\psi}_{\bar{y}\,\bar{y}}^2 \frac{\sigma B_0^2 \bar{x}^{2s}}{\rho C_p}\bar{\psi}_{\bar{y}}^2 \quad (12)$$
$$+\tau\begin{bmatrix} A^{\alpha_4-2\alpha_2+\alpha_5}D_B\left((\bar{T}_w-\bar{T}_\infty)\bar{\theta}\right)_{\bar{y}}\left((\bar{C}_w-\bar{C}_\infty)\bar{\varphi}\right)_{\bar{y}} \\ +A^{2\alpha_4-2\alpha_2}\frac{D_T}{T_\infty}\left(\left((\bar{T}_w-\bar{T}_\infty)\bar{\theta}\right)_{\bar{y}}\right)^2 \end{bmatrix}$$

$$A^{\alpha_5+\alpha_3-\alpha_2-\alpha_1}\bar{\psi}_{\bar{y}}\left((\bar{C}_w-\bar{C}_\infty)\bar{\varphi}\right)_{\bar{x}} - A^{\alpha_5+\alpha_3-\alpha_2-\alpha_1}$$
$$\bar{\psi}_{\bar{x}}\left((\bar{C}_w-\bar{C}_\infty)\bar{\varphi}\right)_{\bar{y}} = A^{\alpha_5-2\alpha_2}D_B\left((\bar{C}_w-\bar{C}_\infty)\bar{\varphi}\right)_{\bar{y}\,\bar{y}}$$
$$+ A^{\alpha_4-2\alpha_2}\frac{D_T}{T_\infty}\left((\bar{T}_w-\bar{T}_\infty)\bar{\theta}\right)_{\bar{y}\,\bar{y}} + A^{\alpha_5}K_r\bar{\varphi}(\bar{C}_w-\bar{C}_\infty) \quad (13)$$

The boundary conditions are

At $y = 0$: $A^{\alpha_3-\alpha_2}\bar{\psi}_{\bar{y}} = A^{m\alpha_1}C_1\bar{x}^m$, $A^{\alpha_3-\alpha_1}\bar{\psi}_{\bar{x}} =$
$-A^{n\alpha_1}C_2\bar{x}^n$ $A^{\alpha_4}(\bar{T}_w-\bar{T}_\infty) = A^{r\alpha_1}C_3\bar{x}^r$, $A^{\alpha_5}(\bar{C}_w-\bar{C}_\infty) = A^{r\alpha_1}C_4\bar{x}^r$
(14.1)

At $y = \infty$: $A^{\alpha_3-\alpha_2}\bar{\psi}_{\bar{y}} = 0$, $(T-T_\infty) =$
$A^{\alpha_4}(\bar{T}-\bar{T}_\infty) = 0$, $(C-C_\infty) = A^{\alpha_5}(\bar{C}-\bar{C}_\infty) = 0$
(14.2)

Invariance of the basic equations yields,
$$\alpha_1 = \frac{2}{1-m}\alpha_2, \quad \alpha_3 = \left(\frac{1+m}{1-m}\right)\alpha_2, \quad s = \frac{m-1}{2},$$
$$n = \frac{m-1}{2}, \quad \alpha_4 = \alpha_5 = r\alpha_1 \quad \text{where } r = \frac{m+1}{2} \quad (15)$$

This yields the invariant transformations group,

$$G: \begin{cases} x = A^{\frac{2}{1-m}}\bar{x}; \quad y = A^{\alpha_2}\bar{y}; \\ \psi = A^{\left(\frac{1+m}{1-m}\right)}\bar{\psi}; \quad \theta = \bar{\theta} \\ \varphi = \bar{\varphi}; \quad (T-T_\infty) = A^{\left(\frac{1+m}{1-m}\right)}(\bar{T}-\bar{T}_\infty); \\ (C-C_\infty) = A^{\left(\frac{1+m}{1-m}\right)}(\bar{C}-\bar{C}_\infty) \end{cases} \quad (16)$$

Now to determine an absolute invariant, namely $\zeta$ and $\zeta = y^N$. For this we write,

$$G: \begin{cases} x = B\bar{x}; \quad y = B^{\left(\frac{1-m}{2}\right)}\bar{y}; \quad \psi = B^{\left(\frac{1+m}{2}\right)}\bar{\psi}; \\ \theta = \bar{\theta}; \quad \varphi = \bar{\varphi}; \quad (T-T_\infty) = B^{-\left(\frac{1+m}{2}\right)}(\bar{T}-\bar{T}_\infty); \\ (C-C_\infty) = B^{-\left(\frac{1+m}{2}\right)}(\bar{C}-\bar{C}_\infty) \text{ where } B = A^{\left(\frac{2}{1-m}\right)} \end{cases} \quad (17)$$

To establish the relation, $y\,x^N = \bar{y}\,\bar{x}^N$ we have $y\,x^N = B^{\left(\frac{1-m}{2}\right)+N}\bar{y}\,\bar{x}^N$.

Putting $\left(\frac{1-m}{2}\right) + N = 0$ we get $y\,x^N = \bar{y}\,\bar{x}^N$.

Since $N = \left(\frac{m-1}{2}\right)$ and $\zeta = y\,x^{\left(\frac{m-1}{2}\right)}$ which is absolute invariant.

Now to find the second absolute invariant $f(\zeta)$ containing dependent variable $\psi$ we assume that $f(\zeta) = \bar{x}^L\bar{\psi}$.

Now $x^L\psi = B^{\left(\frac{1+m}{2}\right)+L}\bar{x}^L\bar{\psi}$

Substituting $\left(\frac{1+m}{2}\right) + L = 0$ we get $L = -\left(\frac{1+m}{2}\right)$.

Thus the absolute invariant $f(\zeta)$ becomes

$$f(\zeta) = x^{-\left(\frac{1+m}{2}\right)}\psi, \text{ finally } \psi = x^{\left(\frac{1+m}{2}\right)}f(\zeta)$$

Similarly, we obtained $T_w - T_\infty = C_3\,x^{-\left(\frac{1+m}{2}\right)}$

and $C_w - C_\infty = C_4\,x^{-\left(\frac{1+m}{2}\right)}$.

We also have $\theta = \theta(\zeta)$ and $\varphi = \varphi(\zeta)$.

From equation (15), the similarity solution exists provided, the sheet is tighten with a speed $U(x) = C_1 x^m$

$$B(x) = B_0\,x^{\left(\frac{m-1}{2}\right)}, \quad v_w(x) = C_2\,x^{\left(\frac{m-1}{2}\right)} \quad (18)$$

Thus the similarity and dependent variables takes the form

$$\zeta = y\,x^{\left(\frac{m-1}{2}\right)}, \quad \psi = x^{\left(\frac{1+m}{2}\right)}f(\zeta), \quad \theta = \theta(\zeta),$$
$$\varphi = \varphi(\zeta) \quad T_w - T_\infty = C_3\,x^{-\left(\frac{1+m}{2}\right)}, \quad (19)$$
$$C_w - C_\infty = C_4\,x^{-\left(\frac{1+m}{2}\right)}$$

To ignore the fluid properties we considered the following transformations

$$\eta = y\sqrt{\frac{m+1}{2}\frac{U(x)}{\nu x}}, \quad \psi = \sqrt{\frac{2}{m+1}\nu x\,U(x)}\,f(\eta) \quad (20)$$

$$\theta = \theta(\eta) \qquad (21)$$

$$\phi = \phi(\eta) \qquad (22)$$

$$T_w - T_\infty = C_3\, x^{-\left(\frac{1+m}{2}\right)}, \quad C_w - C_\infty = C_4\, x^{-\left(\frac{1+m}{2}\right)} \qquad (23)$$

Substituting equations (20)-(23) into the equations (6)-(8) with boundary conditions (9.1, 2), we have

$$f''' + ff'' - \left(\frac{2m}{m+1}\right) f'^2 - \left(\frac{2}{m+1}\right) Mf' = 0 \qquad (24)$$

$$\theta'' + \Pr\theta f' + \Pr f\theta' + \Pr Ec\, f''^2 + \left(\frac{2}{m+1}\right) \Pr M\, Ec\, f'^2 \qquad (25)$$
$$+ \Pr Nb\, \theta'\varphi' + \Pr Nt\, \theta'^2 = 0$$

$$\varphi'' + Le\, f'\varphi + Le\, f\varphi' + \left(\frac{Nt}{Nb}\right)\theta'' + \left(\frac{2}{m+1}\right)\gamma Le\, \mathrm{Re}_x\, \varphi = 0 \qquad (26)$$

The boundary conditions are

$$f(0) = f_w,\quad f'(0) = 1,\quad \theta(0) = 1,\quad \varphi(0) = 1 \qquad (27.1)$$

$$f'(\infty) = 0,\quad \theta(\infty) = 0,\quad \varphi(\infty) = 0 \qquad (27.2)$$

were prime stands for differentiation with respect to $\eta$, $M = \left(\frac{\sigma B_0^2}{\rho C_1}\right)$ is magnetic parameter, $\Pr = \left(\frac{\nu}{\alpha}\right)$ is Prandtl number, $Ec = \left(\frac{U^2}{C_p(T_w - T_\infty)}\right)$ is Eckert number, $Le = \left(\frac{\nu}{D_B}\right)$ is Lewis number, $\mathrm{Re}_x = \left(\frac{xU(x)}{\nu}\right)$ is local Reynolds number, $\gamma = \left(\frac{\nu K_r}{U^2}\right)$ is chemical reaction parameter, $Nb = \left(\frac{(\rho C)_p D_B (C_w - C_\infty)}{(\rho C)_f \nu}\right)$ is Brownian motion parameter, $Nt = \left(\frac{(\rho C)_p D_T (T_w - T_\infty)}{(\rho C)_f T_\infty \nu}\right)$ is Thermophoresis parameter, $\tau = \frac{(\rho C)_p}{(\rho C)_f}$ is the ratio of the nanoparticle heat capacity with base fluid heat capacity, $f_w = -C_2 \sqrt{\frac{2}{(m+1)\nu C_1}}$ is the parameter of suction/injection.

The physical quantities like local skin friction coefficient $C_f$, local Nusselt number $Nu$ and local Sherwood number $Sh$ are defined by

$$C_f = \frac{\tau_w}{\left(\frac{\rho U^2(x)}{2}\right)},\quad Nu = \frac{xq_w}{k(T_w - T_\infty)} \qquad (28)$$

$$\text{and}\quad Sh = \frac{xq_m}{D_B(C_w - C_\infty)}$$

Where the surface shear stress $\tau_w$, heat flux $q_w$ and the mass flux $q_m$ are

$$\tau_w = \mu \frac{\partial u}{\partial y}(x,0),\quad q_w = -k \frac{\partial T}{\partial y}(x,0) \text{ and}$$

$$q_m = -D_B \frac{\partial C}{\partial y}(x,0) \qquad (29)$$

Using the quantities of equation (29) into (28), the reduced Nusselt number $Nur$ and reduced Sherwood number $Shr$ can be obtain in terms of dimensionless temperature and concentration at the surface of sheet respectively as

$$Nur = \sqrt{\frac{2}{m+1}}\, \mathrm{Re}_x^{-1/2} Nu = -\theta'(0) \qquad (30)$$

$$Shr = \sqrt{\frac{2}{m+1}}\, \mathrm{Re}_x^{-1/2} Sh = -\varphi'(0) \qquad (31)$$

And $C_f$ which is the local skin friction coefficient given by

$$\frac{1}{2}\sqrt{\frac{2}{m+1}}\, \mathrm{Re}_x^{1/2} C_f = f''(0) \qquad (32)$$

It is appropriate to mention that neglecting effects of magnetic field strength, equation (24) along with the boundary conditions (27.1, 2) has the analytical solution $f(\eta) = 1 - e^{-\eta}$ discovered by Crane [6].

### IV. NUMERICAL TREATMENT

The reduced system of equations (24)-(26) along with the boundary conditions (27.1, 2) gives non-linear coupled BVP of second and third order. An efficient Runge-Kutta fourth order integration technique with a guessing of $f''(0)$, $\theta'(0)$ and $\varphi'(0)$ values by Newton-Raphson shooting scheme till the boundary conditions at $f''(\infty)$, $\theta'(\infty)$ and $\varphi'(\infty)$ decline rapidly to zero, has been employed numerically. The present method involves transformation of the governing system of equations to a system of seven coexisting equations of first order for seven unknowns following the method of superposition described by Na (1982). To handle these simultaneous equations we required the same number of initial conditions. Here we have two initial conditions in $f$ and one in each of $\theta$ and $\varphi$ are known. Thus we are short of three conditions $f''(0)$, $\theta'(0)$ and $\varphi'(0)$ that can be generated by using the shooting method and employing the values of $f''(\infty)$, $\theta'(\infty)$ and $\varphi'(\infty)$. The step size $\Delta \eta = 0.001$ is used to find the numerical solution with $\eta_\infty$ and accuracy to the four decimal places is considered. To evaluate $\eta_\infty$ we begin with guess value and solve BVP for a specific set of parameters to obtain $f''(0)$, $\theta'(0)$ and $\varphi'(0)$. The series of actions is repeated for big values of $\eta_\infty$ till two successive values of $f''(0)$, $\theta'(0)$ and $\varphi'(0)$ vary only after specific digit indicating the limit of the boundary along $\eta$. The obtained final value is considered as the correct value for that set of parameters.

### V. RESULTS AND DISCUSSION

In the previous section the highly nonlinear system of governing equations is solved explicitly for the particular values of associated physical parameters. The velocity profile $f'$, temperature profile $\theta$ and concentration profile $\varphi$ are studied for the different values of associated parameters. Prandtl number $Pr$ of air is chosen as 0.71 at temperature $25^0 C$ and 1 atmospheric pressure. Thermophoresis parameter $Nt$ and the Brownian motion parameter $Nb$ upon nature of flow and transport take value 0.5. Also Lewis number $Le$, the Eckert number $Ec$ and the Chemical reaction parameter $\gamma$ is chosen as 0.5. Suction/injection parameters hold the values $0, \pm 0.3, \pm 0.5$, an index parameter $m$ holds the values 0.3, 1, 3, 4, 5 and that of magnetic parameter hold the values 0, 0.3, 0.9, 1.5 and 2. Comparing the results obtained with the existing methods there is a marked improvement observed by neglecting the effects of physical parameters $Nt$, $Nb$ and $fw = 0$ with the index parameter $m = 1$ (Table 1). The variation among the reduced Nusselt number $Nur$ and the reduced Sherwood number $Shr$ with $0.1 \le Nb \le 0.5$ and $0.1 \le Nt \le 0.5$ for $Pr = 10$ and $Le = 10$ with the index parameter $m = 1$ can be observed (Table 2). It is noted that $Nur$ is a decreasing function, while $Shr$ is an increasing function of each dimensionless parameters considered $Nb$, $Nt$, $Pr$ and $Le$. The proposed method provides reliable results compared to the results published by Makinde and Aziz [15] (Table 2).

Figs.1-2 shows typical velocity profiles for different values of magnetic and the index parameter. Fig.1 reveals the facts that addition in the magnetic field retards the fluid velocity. The velocity significantly increased with an increase in the index parameter m (Ref. Fig.2). The similarity variable $\theta(\eta)$ for selected parameters is plotted. Fig. 3 reveals that, addition in the magnetic parameter results thick temperature boundary layer whereas from Fig. 4, it is seen that the temperature decline with growth in m.

Fig.5-7 depicts the influences of pertinent parameters of the fluid concentration. It revealed that the concentration becomes highest on the sheet surface, but decreases to zero. Fig. 5 depicts the effects of chemical reaction parameter $\gamma$ on the concentration profile. We observed the concentration profiles decline with growing values of $\gamma$. Fig.6 depicts, increase in magnetic parameter M causes the increase in nanoparticle concentration. Thus, the mass transfer rate $-\varphi'(0)$ i.e. local Sherwood number declines with increase in magnetic parameter M. Nusselt and Sherwood number are synonymous to heat and mass transfer rate at the surface of sheet respectively. Fig. 7 reveals that, as m varies, the concentration decreases. Figs. 8 and 9 illustrate the results of the skin-friction coefficient. From Fig. 8, we observed that the local skin-friction coefficient decreases as an increase in the parameter M. Fig. 9 analysed the dimensionless skin friction coefficient with distinct amounts of m. Note the skin friction coefficient reduces with increase in m. Heat transfer rate and $Nt$ parameter variation are plotted in Figs. 10 and 11. From Fig. 10, it is clear that the heat transfer rate is enhanced as the magnetic parameter M raises whereas Fig. 11 illustrates downturn in the heat transfer rate as the m numbers increase. Figs.12-14 depicts the dimensionless mass transfer rates for a chemical reaction parameter $\gamma$, the magnetic parameter M and the index parameter m respectively. From Fig.12 reveals, additions in the value of $\gamma$ decline the rate of mass transfer. Fig. 13 illustrates the effects of m numbers on the rate of mass transfer and it is seen that the mass transfer rates decreases as m numbers increase. Fig. 14 reveals that, as magnetic parameter M increases there is decrease in the mass transfer rate.

| Pr | Makinde and Aziz (2011) | Khan and Pop (2010) | Wang (2009) | Existing Outcome |
|---|---|---|---|---|
| 0.07 | 0.0656 | 0.0663 | 0.0656 | 0.0654 |
| 0.20 | 0.1691 | 0.1691 | 0.1691 | 0.1690 |
| 0.70 | 0.4539 | 0.4539 | 0.4539 | 0.4538 |
| 2.00 | 0.9114 | 0.9113 | 0.9114 | 0.9112 |
| 7.00 | 1.8954 | 1.8954 | 1.8954 | 1.8952 |
| 20.00 | 3.3539 | 3.3539 | 3.3539 | 3.3538 |
| 70.00 | 6.4622 | 6.4621 | 6.4622 | 6.4620 |

Table 1: Comparison of $-\theta'(0)$ with the literature

| Nb | Nt | Nur (2011) | Shr (2011) | Nur Existing | Shr Existing |
|---|---|---|---|---|---|
| 0.1 | 0.1 | 0.9524 | 2.1294 | 0.9522 | 2.1290 |
| 0.2 | 0.1 | 0.5056 | 2.3819 | 0.5054 | 2.3815 |
| 0.3 | 0.1 | 0.2522 | 2.4100 | 0.2520 | 2.4098 |
| 0.4 | 0.1 | 0.1194 | 2.3997 | 0.1192 | 2.3992 |
| 0.5 | 0.1 | 0.0543 | 2.3836 | 0.0540 | 2.3831 |
| 0.1 | 0.2 | 0.6932 | 2.2740 | 0.6929 | 2.2736 |
| 0.1 | 0.3 | 0.5201 | 2.5286 | 0.5200 | 2.5282 |
| 0.1 | 0.4 | 0.4026 | 2.7952 | 0.4024 | 2.7949 |
| 0.1 | 0.5 | 0.3211 | 3.0351 | 0.3209 | 3.0347 |

Table 2: Comparison of $-\theta'(0)$ and $-\varphi'(0)$ with the literature

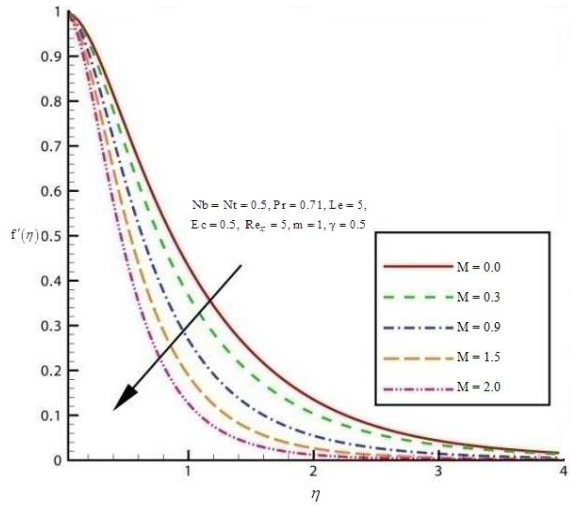

Fig. 1 Impact of M on velocity profile for particular parameters.

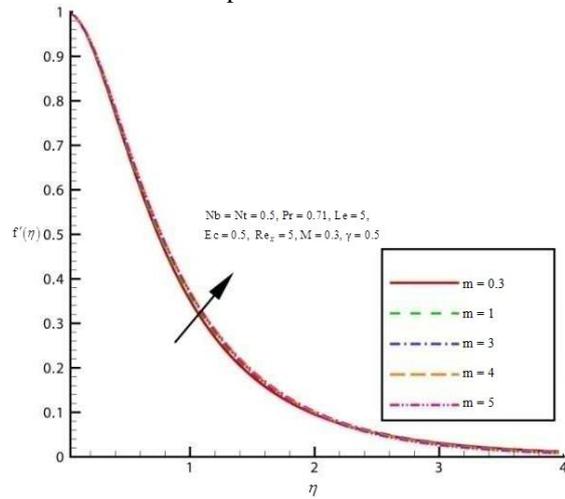

Fig. 2 Impact of m on velocity profile for particular parameters

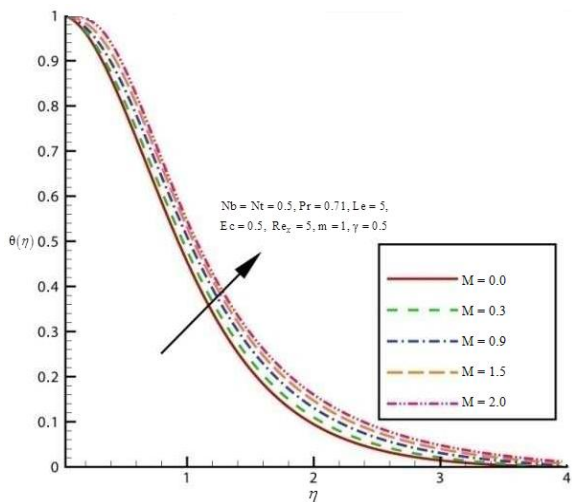

Fig.3. Impact of M on temperature profile for particular parameters.

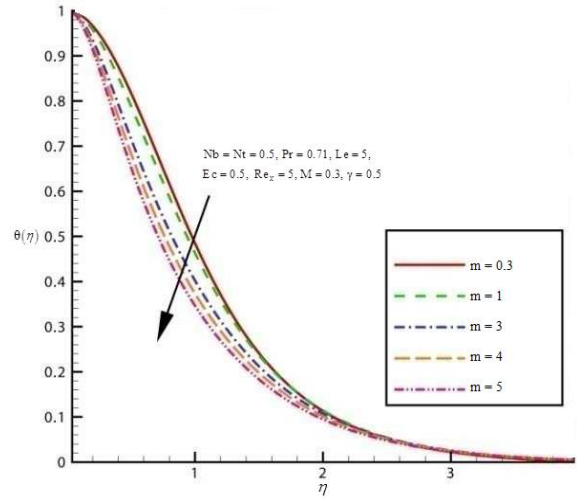

Fig.4. Impact of m on temperature profile for particular parameters.

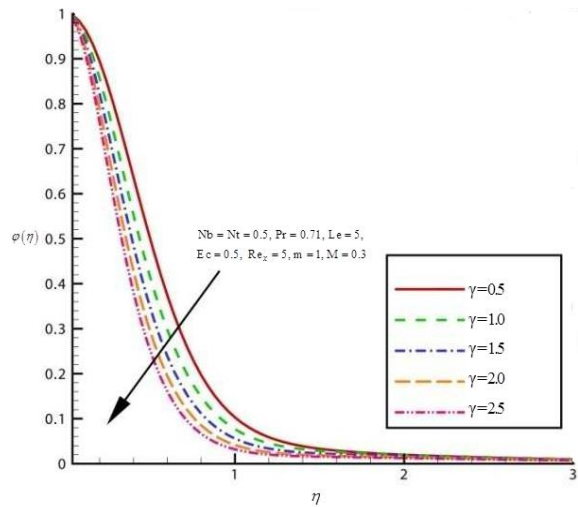

Fig.5. Impact of $\gamma$ on concentration profile for particular parameters.

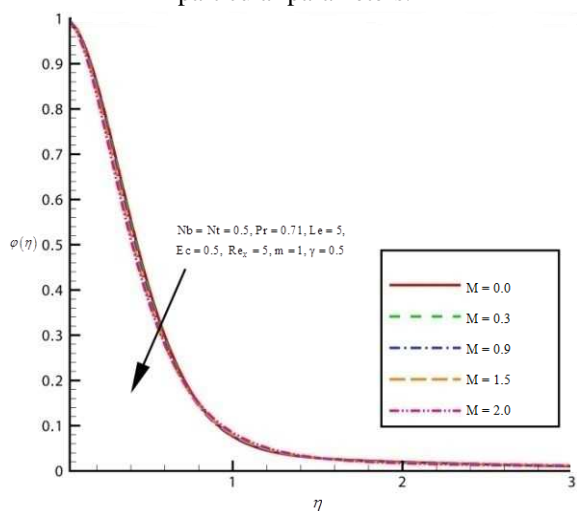

Fig.6. Impact of M on concentration profile for particular parameters.

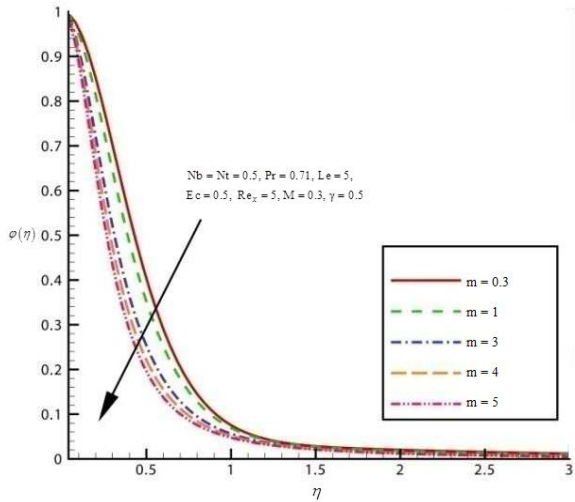

Fig.7. Impact of m on concentration profile for particular parameters.

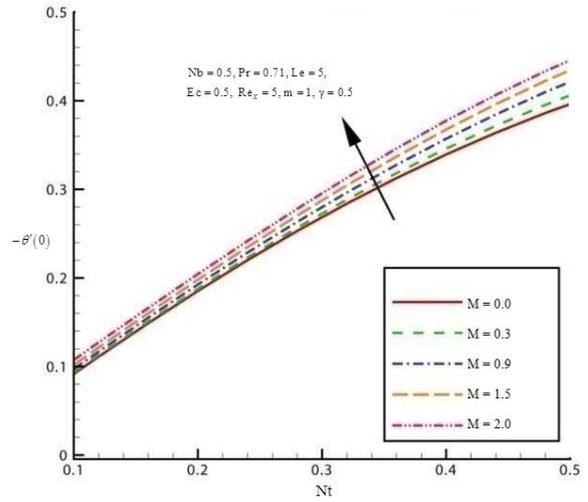

Fig.10. Impact of M on heat transfer rate for particular parameters.

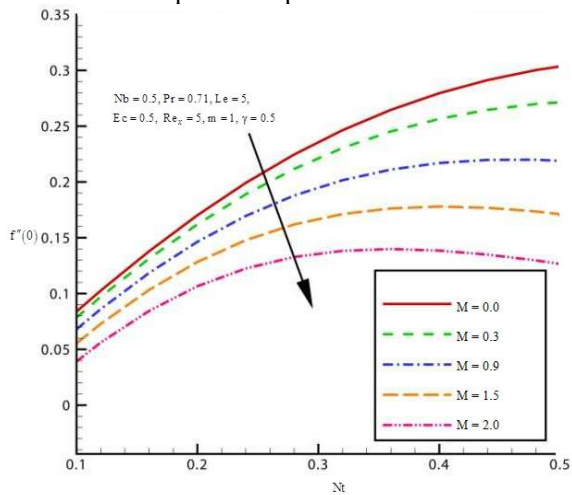

Fig.8. Impact of M on skin friction coefficient for particular parameters.

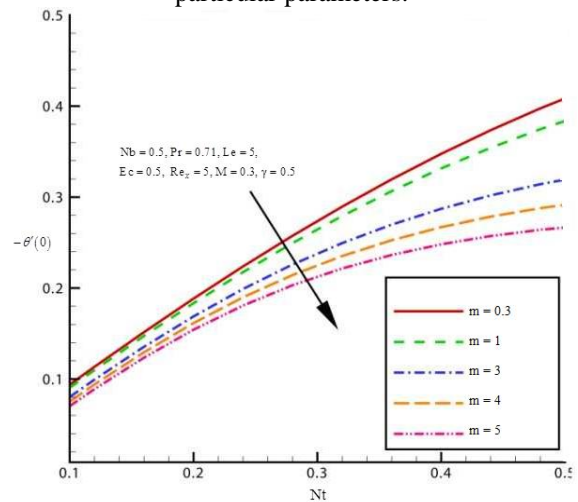

Fig.11. Impact of m on heat transfer rate for particular parameters.

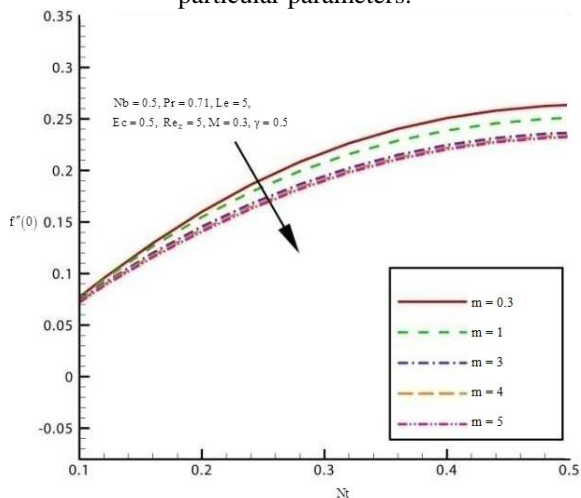

Fig.9. Impact of m on skin friction coefficient for particular parameters.

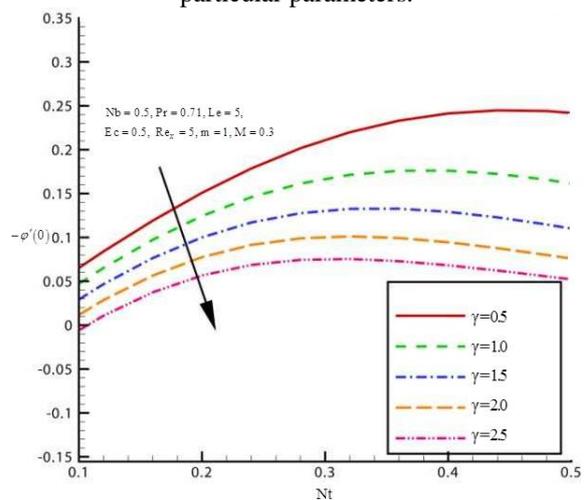

Fig.12. Impact of $\gamma$ on mass transfer rate for particular parameters

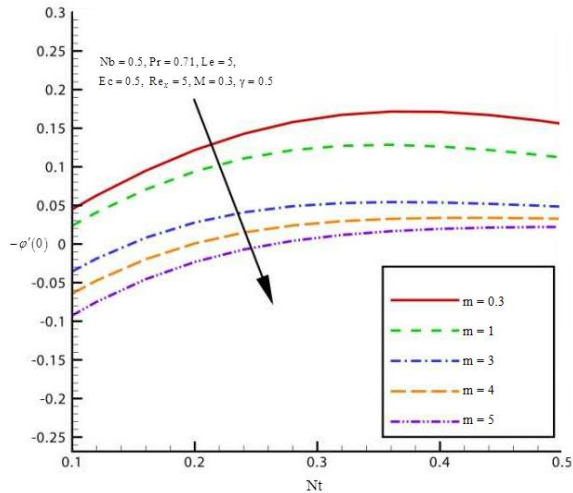

Fig.13. Impact of m on mass transfer rate for particular parameters.

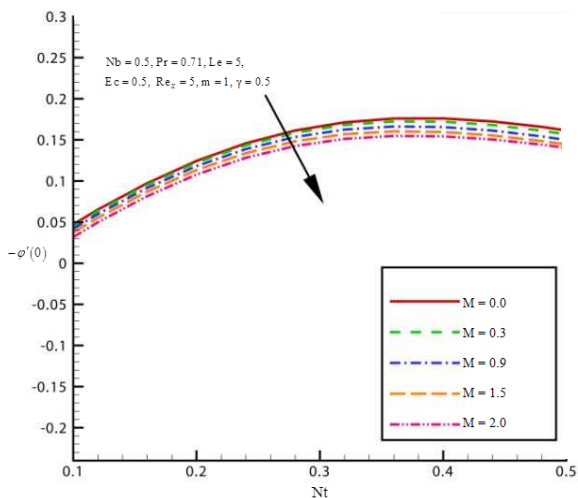

Fig.14. Impact of M on mass transfer rate for particular parameters.

### VI. CONCLUSIONS

The research presented here analyzed the problem of steady MHD free convective heat and mass transfer past a stretching sheet in existence of chemical reaction for the flow of viscous incompressible Nanofluid with suction or injection. The governing PDEs are transformed into non-linear ODEs using group theoretic approach and solved using explicit numerical techniques. Numerical outcomes for the dimensionless parameters and the local skin friction coefficient, the reduced Nusselt number and the reduced Sherwood number are studied graphically and analysed quantitatively. It is observed that $Nu_r$ is a depressing whereas $Sh_r$ is a growing function of each dimensionless parameters $Nb$, $Nt$ Pr and $Le$. The analysis presented here shows that a further increase to the magnetic parameter gives additional reductions in velocity and the concentration profile and increases the temperature profile. The velocity profile increases with increase in size of the index parameter m, whereas the temperature and the concentration profiles significantly decreased. Advancement in the chemical reaction parameter causes reduction in the concentration profile and the mass transfer rate with rising values of magnetic parameter. Furthermore, addition in index parameter m causes significant degradation of heat and mass transfer rate at the surface of stretching sheet and the skin-friction coefficient.

### REFERENCES

[1] Afify A.A (2004), MHD free convective flow and mass transfer over a stretching sheet with chemical reaction, *Heat and Mass Transfer*, **40** (6-7) 495-500. http://dx.doi.org/10.1007/s00231-003-0486-0

[2] Buongiorno J (2006). Convective Transport in Nanofluids, *ASME Journal of Heat Transfer*, *128(3)*, 240-250. http://dx.doi.org/10.1115/1.2150834.

[3] Chaim T.C (1995). Hydromagnetic Flow over a Surface with a Power-Law Velocity, *Int. J Eng. Sci., 33 (3,)* 429-435. http://dx.doi.org/10.1016/0020- 7225(94)00066-S

[4] Chamkha A.J (2003). MHD flow of a uniformly stretched vertical permeable surface in the presence of heat generation/absorption and a chemical reaction, *Int. Comm. Heat Mass Transfer,30, 413*-422. http://dx.doi.org/10.1016/S0735-1933(03)00059-9

[5] Choi.S.U.S. (1995). Enhancing Thermal Conductivity of Fluids with Nanoparticles, Developments and Applications of Non-Newtonian Flows, eds. D. A. Siginer and H. P. Wang, A*SME FED-Vol.231/MD-Vol.66*, 99-105.

[6] Crane L.J (1970). Flow past a stretching plate, *ZAMP*, **21** (4) 645-647. http://link.springer.com/journal/33

[7] Das S.K, Choi S.U.S, Yu. W, Pradeep T. (2007). Nanofluids: *Science and Technology, Wiley*, New Jersey, http://dx.doi.org/10.1002.

[8] Eastman J., Choi S.U.S, Lib S., Yu. W, Thompson L.J. (2001). Anomalously Increased Effective Thermal Conductivities of Ethylene-Glycol-Based Nanofluids Containing Copper Nanoparticles, *Applied Physics Letters, 78* 718-720 http://dx.doi.org/10.1063/1.1341218.

[9] Gorla R.S.R, Chamkha A (2011). Natural Convective Boundary Layer Flow over a Horizontal Plate Embedded in a Porous Medium Saturated with a Nanofluid, *Journal of Modern Physics*, **2**, 62-71. http://dx.doi.org/10.43236/jmp.2011.22011

[10] Gorla R.S.R, El-Kabeir S.M.M, Rashad A.M (2011). Heat Transfer in the Boundary Layer on a Stretching Circular Cylinder in a Nanofluid, *AIAA Journal of Thermo physics and Heat transfer, 25*,183-186. http://dx.doi.org/10.2514/1.51615

[11] Kakac S, Pramuanjaroenkij, A. (2009). J (2006). Review of Convective Convective Heat Transfer Enhancement with Nanofluids, *Int. J. Heat Mass Transfer, 52,* 3187-3196. http://dx.doi.org/10.1016/j.ijheatmasstransfer.2009.02.006.

[12] Khan W.A, Pop I (2010). Boundary-layer flow of a nanofluid past a stretching sheet, *Int. J. Heat Mass Transf.,* **53**, 2477-2483. http://dx.doi.org/10.1016/j.ijheatmasstransfer.2010.01.032

[13] Kuznetsov A.V, Nield D.A (2014). Natural convective boundary-layer flow of a nanofluid past a vertical plate: A revised model, *Int. J. Therm. Sci.,* **77**,126-129. http://dx.doi.org/10.1016/j.ijthermalsci.2013.10.007.

[14] Kuznetsov A.V, Nield D.A (2010). Natural convective boundary- layer flow of a nanofluid past a vertical plate, *Int. J. Therm. Sci., 49 (2)*, 243-247.

[15] Makinde O.D, Aziz A (2011) Boundary layer flow of a nanofluid past a stretching sheet with a convective boundary condition**,** *Int. J. Therm. Sci.,50 (7),*1326-1332. http://dx.doi.org/10.1016/j.ijthermalsci.2011.02.019

[16] Na T.Y (1982). Computational methods in engineering boundary value problems, New York: Academic Press.

[17] Nield. D.A, Bejan A, (2013). Convection in porous media (4<sup>th</sup> Edition), Springer, New York, http://dx.doi.org/ 10.1007/978-1-4614-5541-7
[18] Palanimani P.G (2007), Effects of chemical reactions, heat, and mass transfer on nonlinear magneto-hydrodynamic boundary layer flow over a wedge with a porous medium in the presence of ohmic heating and viscous dissipation, Journal of Porous Media, **10**(5) 489-502. http://dx.doi.org/10.1615/JPorMedia.v10.i5.60
[19] Postelnicu A (2007)., Influence of chemical reaction on heat and mass transfer by natural convection from vertical surfaces in porous media considering Soret and Dufour effects, *Heat and Mass Transfer* **43** 595-602. http://dx.doi.org/10.1007/s10483-010-1302-9
[20] Wang L, Wei X. (2009) Heat Conduction in Nanofluids, *Chaos Solutions Fractals,* **39** (2009)2211-2215. http://dx.doi.org/10.1016/j.chaos.2007.06.072.
[21] Yu. D.M., Routbort, J.L, Choi S.U.S. (2008) Review and Comparison of Nanofluid Thermal Conductivity and Heat Transfer Enhancements, *Heat Transfer Engineering,* **29** *(5)* 432-460.